# A Blueprint for Auditing Generative AI

Jakob Mökander,[1,2] Justin Curl,[3] Mihir Kshirsagar[4]

[1] Oxford Internet Institute, University of Oxford, 1 St Giles', Oxford, OX1 3JS, UK
[2] Digital Ethics Center, Yale University, 85 Trumbull St., New Haven, CT 06511, USA
[3] Harvard Law School, 1585 Massachusetts Ave, Cambridge, MA 02138, US
[4] Center for Information Technology Policy, Princeton University, Princeton, NJ 08544, US

## Abstract

The widespread use of generative AI systems is coupled with significant ethical and social challenges. As a result, policymakers, academic researchers, and social advocacy groups have all called for such systems to be audited. However, existing auditing procedures fail to address the governance challenges posed by generative AI systems, which display emergent capabilities and are adaptable to a wide range of downstream tasks. In this chapter, we address that gap by outlining a novel blueprint for how to audit such systems. Specifically, we propose a three-layered approach, whereby governance audits (of technology providers that design and disseminate generative AI systems), model audits (of generative AI systems after pre-training but prior to their release), and application audits (of applications based on top of generative AI systems) complement and inform each other. We show how audits on these three levels, when conducted in a structured and coordinated manner, can be a feasible and effective mechanism for identifying and managing some of the ethical and social risks posed by generative AI systems. That said, it is important to remain realistic about what auditing can reasonably be expected to achieve. For this reason, the chapter also discusses the limitations not only of our three-layered approach but also of the prospect of auditing generative AI systems at all. Ultimately, this chapter seeks to expand the methodological toolkit available to technology providers and policymakers who wish to analyse and evaluate generative AI systems from technical, ethical, and legal perspectives.

# 1 Introduction

The promises and perils of artificial intelligence (AI) have been widely discussed, with generative AI systems being the most recent technological advance to capture the world's attention. Each year, Stanford University publishes an AI index report covering promising trends in research and applications related to AI. The 2023 report describes 2022 as the year that generative AI broke into the public consciousness with popular applications like ChatGPT, Stable Diffusion and Make-A-Video.[1] Amid growing private sector investment and adoption in industries as diverse as healthcare, financial services and cybersecurity, generative AI is already changing how we live and work. Stanford's report gives two examples that help illustrate exactly how this change is occurring. First, DeepMind released a generative AI model that assists scientists researching de novo antibody discovery, accelerating medical research by enabling them to generate and assess more promising candidates more quickly. Second, GitHub released a text-to-code generative tool called Copilot, which reportedly improved users' coding productivity and enjoyment. The model handles the more rote parts of software development, freeing up users to focus on the interesting parts of projects.

Yet one need only read news of AI scams and misuse to see that the technology also creates real societal harms. QTCinderella is one of many popular female Twitch streamers experiencing sexual harassment as AI-generated, non-consensual pornographic images and videos of her are circulated online. These images and videos were created with generative AI tools and sold on a deepfake pornography website without her knowledge.[2] Scammers have also been using generative AI tools to imitate the voices of distressed family members. These scammers have swindled thousands of people out of millions of dollars, leaving affected families with little recourse.[3]

With the aim of addressing these harms, experts from academia, industry and government have called for generative AI systems to be audited. Simplified, auditing in this context refers to structured and independent assessments of how well an AI system's design, properties, or impact adheres to applicable standards, regulations, and norms. In Section 2, we will explain and discuss what AI auditing is in greater detail. For now, the key point to stress is that AI auditing has recently attracted much attention as a promising AI governance mechanism. For example, in their seminal paper on foundation models (which are AI systems adaptable to a wide range of downstream tasks[4]), Bommasani et al. suggested that "such models should be subject to rigorous testing and *auditing* procedures."[5] Similarly, OpenAI's CEO Sam

---

[1] N. Maslej et al., "The AI Index 2023 Annual Report," *AI Index Steering Committee, Institute for Human-Centered AI, Stanford University, Stanford, CA*, 2023.
[2] S. Cole et al., "'You Feel So Violated': Streamer QTCinderella Is Speaking Out Against Deepfake Porn Harassment," *Vice* (blog), February 13, 2023, https://www.vice.com/en/article/z34pq3/deepfake-qtcinderella-atrioc.
[3] P. Verma, "They Thought Loved Ones Were Calling for Help. It Was an AI Scam.," *Washington Post*, March 10, 2023, https://www.washingtonpost.com/technology/2023/03/05/ai-voice-scam/.
[4] We discuss the definition, characteristics and governance of foundation models further in Section 2, highlighting how that discussion applies to generative AI systems.
[5] R. Bommasani et al., "On the Opportunities and Risks of Foundation Models" (arXiv, July 12, 2022), https://doi.org/10.48550/arXiv.2108.07258.

Altman has stated that "it's important that efforts like ours submit to *independent audits* before releasing new systems."[6] Meanwhile, the European Commission has classified generative AI as a "high-risk",[7] thus requiring providers of such systems to undergo conformity assessments (audits by another name).

Against this backdrop, the main contribution offered in this chapter is a novel blueprint for how to conduct generative AI audits. To be clear, our primary aim is not to argue that generative AI systems *should* be audited. Rather, we address the following question: *if* policymakers and technology providers have an appetite to audit generative AI systems, *how* could such audits be structured to be rigorous, feasible and effective in practice? Building on previous work,[8] we propose a three-layered approach, whereby *governance audits* (of technology providers that design and disseminate generative AI systems), *model audits* (of generative AI models after pre-training but prior to their release), and *application audits* (of applications based on generative AI) complement and inform each other. Figure 1 below provides an overview of this three-layered approach.

*Figure 1. Blueprint for how to audit generative AI: A three-layered approach.*

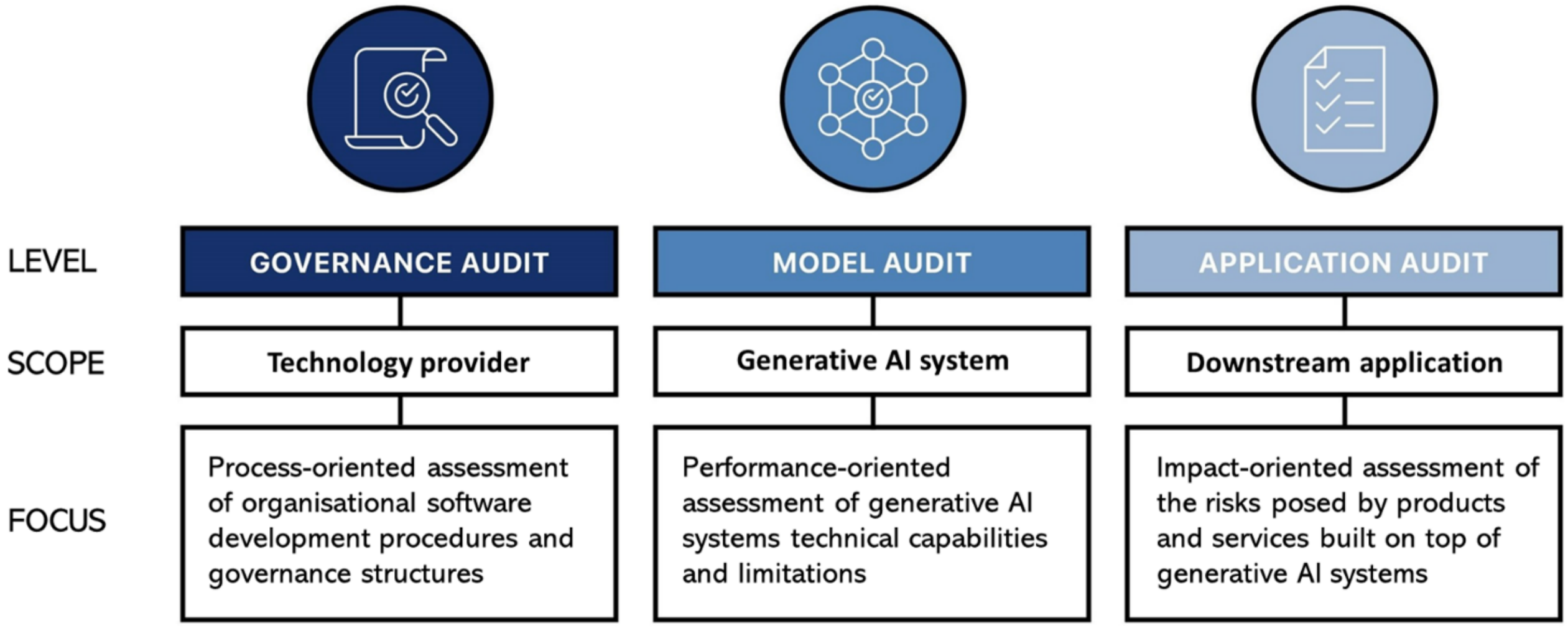

As we will demonstrate throughout this chapter, many tools and methods already exist to conduct governance, model and application audits. However, to provide meaningful assurance for generative AI systems, audits conducted on each of the three levels must be combined into a structured and coordinated procedure. To the best of our knowledge, our blueprint for how to audit generative AI systems is the first of its kind, and we hope it will inform both technology providers' and policymakers' efforts to ensure that generative AI systems are legal, ethical and technically robust.

---

[6] S. Altman, "Planning for AGI and Beyond," 2023, https://openai.com/blog/planning-for-agi-and-beyond.
[7] It is still uncertain how the EU AI Act should be interpreted with respect to generative AI systems. N. Helberger and N. Diakopoulos, "ChatGPT and the AI Act," *Internet Policy Review* 12, no. 1 (2023), https://doi.org/10.14763/2023.1.1682.
[8] J. Mökander et al. "Auditing Large Language Models: A Three-Layered Approach," *AI and Ethics*, May 30, 2023, https://doi.org/10.1007/s43681-023-00289-2.

Before proceeding, we find it important to clarify this chapter's scope. Our three-layered approach concerns the procedure of generative AI audits and answers questions about *what* should be audited, *when,* and according to *which criteria*. When designing a holistic auditing ecosystem, one must consider additional factors like *who* should conduct the audit and *how* to ensure post-audit action. While these considerations are important, they fall outside the scope of this chapter. Designing an institutional audit ecosystem is a non-trivial question that we have neither the space nor the capacity to address here. That formation process will likely be gradual and involve negotiations between AI labs, policymakers and civil rights groups, negotiations we hope our auditing blueprint can help spark and inform.

The remainder of this chapter proceeds as follows. Section 2 defines key terminology and explains why governments, academic researchers and social advocacy groups have recently called for generative AI systems to be audited. Section 3 introduces our blueprint for how to audit generative AI. Specifically, a three-layered approach is proposed, whereby governance, model and application audits complement and reinforce each other. Sections 4–6 detail why each of these three types of audits are needed, what they entail and the outputs they should produce. Section 7 highlights the limitations of our three-layered approach and proposes avenues for further research. Finally, Section 8 concludes by discussing the implications of our findings for technology providers, policymakers and independent auditors.

# 2 Why audit generative AI systems?

Auditing is a systematic process of gathering and evaluating evidence about an entity and communicating the results to relevant stakeholders. Just as audits can be used to create transparency and legal accountability in areas like financial accounting and worker safety, auditing can also be used as a governance mechanism for AI systems. Three ideas underpin the promise of auditing as an AI governance mechanism, namely, that:

i. procedural regularity and transparency contribute to good governance;
ii. proactivity in the design of AI systems helps identify risks and prevent harm before it occurs; and,
iii. the operational independence between the auditor and the auditee contributes to the objectivity of the evaluation.

This field of AI auditing is still relatively new, dating back to Sandvig et al.'s 2014 article *Auditing Algorithms.*[9] Since then, AI auditing has been defined in many ways for many purposes. In some cases, policymakers mandate audits to ensure AI systems *meet specific legal standards.*[10] In other cases, technology firms commission audits to *identify and mitigate risks* in their AI systems. In still others, organisations conduct audits to *inform citizens* about the conduct of specific companies. While AI auditing may be widespread in practice, many of its procedures have been narrowly tailored, focusing on specific sectors or stages of the AI development lifecycle. Kazim et al., for example, created an auditing procedure for AI systems used in the specific context of the hiring process.[11] Raji et al. developed an internal auditing procedure focused on organisational practices, seeking to identify and mitigate risks before a model is deployed.[12] Yet AI capabilities are becoming increasingly general. As they do so, the need for new, more flexible auditing procedures will only grow. To illustrate why the existing, narrow auditing procedures are inadequate, we discuss state-of-the-art generative AI models and the characteristics that make them uniquely difficult to audit.

[9] C. Sandvig et al., "Auditing Algorithms," in *ICA 2014 Data and Discrimination Preconference*, 2014, 1–23, https://doi.org/10.1109/DEXA.2009.55.
[10] Gibson Dunn, "New York City Proposes Rules to Clarify Upcoming Artificial Intelligence Law for Employers," 2023, https://www.gibsondunn.com/new-york-city-proposes-rules-to-clarify-upcoming-artificial-intelligence-law-for-employers/.

Generative AI is a functional definition that describes systems that can output content, such as images, text, audio or code.[13] These models can generate outputs based on a nearly limitless number of potentially harmful inputs. Bommasani et al. coined the overlapping but distinct term *foundation models* to describe models upon which other AI applications and systems are built.[5] These models are effective across a wide range of downstream tasks and can display emergent capabilities. Additionally, whereas advanced AI models were previously designed and deployed by a single actor for a single use case, foundation models tend to be developed by one actor and subsequently adapted by many other actors for many more applications. What makes auditing especially challenging is that the most advanced generative AI systems exhibit characteristics of both generative AI models and foundation models.

With generative AI systems, the nearly limitless output possibilities make comprehensively protecting against risks ahead of time infeasible. With foundation models, compared to past methods, addressing a model's potential harms independent of the context in which it is deployed is more difficult. When those harms do occur, the new AI development process complicates the allocation of responsibility between technology providers and downstream developers. Taken together, the capabilities and training processes of generative AI foundation models have outpaced the development of the tools and procedures responsible for legal oversight, suggesting we need new approaches to address the potential harms of generative AI.

An article by Weidinger et al. identifies these harms more systematically by focusing on the specific subcategory of AI models that are both foundational and generative: large language models (LLMs).[14] We reference these harms throughout this chapter because i) many of the risks of LLMs generalize to all forms of generative AI and ii) the most advanced models, such as GPT-4, are increasingly trained with multiple types of media (text, images, etc.), blurring the line between the harms of LLMs and generative AI models. Using Weidinger et al.'s typology, the risks and harms associated not only with LLMs but also with generative AI systems can be divided into six categories:

1) *Discrimination*: generative AI systems can create representational and allocative harms by perpetuating social stereotypes and inequities;
2) *Information hazards*: generative AI systems may leak private or sensitive information;
3) *Misinformation hazards:* generative AI systems may produce inaccurate content that misleads users or erodes their trust in information;
4) *Malicious use*: generative AI systems can be co-opted by malicious actors to cause harm (e.g., personalized scams or deepfake revenge porn);
5) *Human-computer interaction harms*: users may overestimate the capabilities of generative AI systems that appear human-like and use them in unsafe ways; and
6) *Automation and environmental harms:* training and operating generative AI systems require lots of computing power, incurring high environmental costs.

[11] E. Kazim et al., "Systematizing Audit in Algorithmic Recruitment," *J Intell* 9, no. 3 (2021): 1–11, https://doi.org/10.3390/jintelligence9030046.
[12] I. D. Raji et al., "Closing the AI Accountability Gap: Defining an End-to-End Framework for Internal Algorithmic Auditing," in *Proceedings of the 2020 Conference on Fairness, Accountability, and Transparency*, FAT* '20 (New York, NY, USA: Association for Computing Machinery, 2020), 33–44, https://doi.org/10.1145/3351095.3372873.
[13] OpenAI, "Generative Models," 2016, https://openai.com/research/generative-models.
[14] L. Weidinger et al., "Ethical and Social Risks of Harm from Language Models" (arXiv, December 8, 2021), https://doi.org/10.48550/arXiv.2112.04359.

Our blueprint for how to audit generative AI systems has been designed to help organisations identify and manage the above-listed risks. However, before outlining our three-layered approach in detail, it is useful to consider some reasonable objections to it and provide tentative responses to these objections. First, one might argue there is no need to audit generative AI per se and these audits should be conducted at the application level instead. Although application-level audits are important, that argument presents a false dichotomy: governance mechanisms can and should be established at different stages of the supply chain. While some risks can only be addressed at the application level, others are best managed upstream (e.g., those related to sourcing training datasets). Many factors determine whether a technological artefact causes harm, but technology providers must still take precautions for reasonably foreseeable risks in the stages of development they control. For this reason, we propose complementing application audits with governance audits of organisations that develop generative AI models.

Second, one may object to auditing because it is impossible to identify and mitigate all the risks of generative AI at the technology level. As we explain in Section 5, this impossibility arises in part because people legitimately disagree about how a technology should be used and what constitutes harm. Broad terms like "fairness" and "transparency" can hide these deep normative disagreements, but they exist nevertheless.[15] For example, collecting individual data to improve facial recognition systems may reduce racial bias in these algorithms but comes at the cost of reduced privacy. As another example, fairness can be statistically defined in many ways (like demographic parity and counterfactual fairness), yet these definitions are often mutually exclusive.[16] Audits alone cannot guarantee that systems are "ethical" in a universal sense, but they still contribute to good governance in several ways. They can help technology providers identify risks and prevent potential harm. They can inform the continuous (re-design) of generative AI systems. And perhaps most importantly, they can make implicit choices and tensions visible, allowing for policy solutions that – even when imperfect – are clearer and better understood.

[15] Watson & Mökander (2023). "In Defense of Sociotechnical Pragmatism." (The 2022 Yearbook of the Digital Ethics Lab).
[16] S. Barocas, M. Hardt, and A. Narayanan, *Fairness and Machine Learning: Limitations and Opportunities* (2023).

Third, one may object that designing auditing procedures for generative AI is impractically difficult. We agree that designing these procedures includes both practical and conceptual challenges. Practically, different stages in the software development life cycle (e.g., curating training data and the pre-training/fine-tuning of model weights) overlap in messy ways. That open-source systems are continuously re-trained and re-uploaded on platforms like Hugging Face makes it difficult to know which model versions to audit and when.[17] Conceptually, the challenges run even deeper. Definitions of disinformation and hate speech are contested, and even seemingly straightforward notions like "truthfulness" and "fairness" are hard to operationalise universally. This makes establishing a baseline for assessing generative AI systems especially difficult.

Importantly, these difficulties are not reasons for abstaining from developing auditing procedures for generative AI. Instead, they remind us that no single method can cover all generative AI-related risks in all contexts. However, the fact that audits might be insufficient does not mean they cannot be a useful complement to other governance mechanisms.

# 3 How to audit generative AI systems?

AI audits can be structured in many ways, and a wide range of AI auditing procedures have already been developed.[18] However, not all auditing procedures are equally effective in handling the risks posed by generative AI systems. Nor are they all equally likely to be implemented, due to factors including technical limitations, institutional access and administrative costs. In this section, we first introduce our blueprint for how to audit generative AI systems. The question thus remains: how can auditing procedures for generative AI systems be designed that are feasible and effective in practice?

In this section, we build on previous research in the fields of IT audits and systems engineering to outline a blueprint for how to audit generative AI systems. Specifically, we propose a three-layered approach. First, technology providers developing generative AI systems should undergo *governance audits* that assess their organisational procedures, accountability structures and quality management systems. Second, generative AI models should undergo *model audits*, assessing their capabilities and limitations after initial training but before adaptation and deployment in specific applications. Third, downstream applications using generative AI models should undergo continuous *application audits* that assess the ethical alignment and legal compliance of their intended functions and their impact over time. Figure 2 illustrates the details of our three-layered approach.

[17] H. R. Kirk et al., "Bias Out-of-the-Box: An Empirical Analysis of Intersectional Occupational Biases in Popular Generative Language Models," in *NeurIPS*, 2021, https://github.com/oxai/intersectional_gpt2.
[18] J. Mökander. "Auditing of AI: Legal, Ethical, and Technical Approaches." *Digital Society*. 2023. https://doi.org/10.1007/s44206-023-00074-y

*Figure 2. Governance, model and application audits complement each other in terms of scope and methodology.*

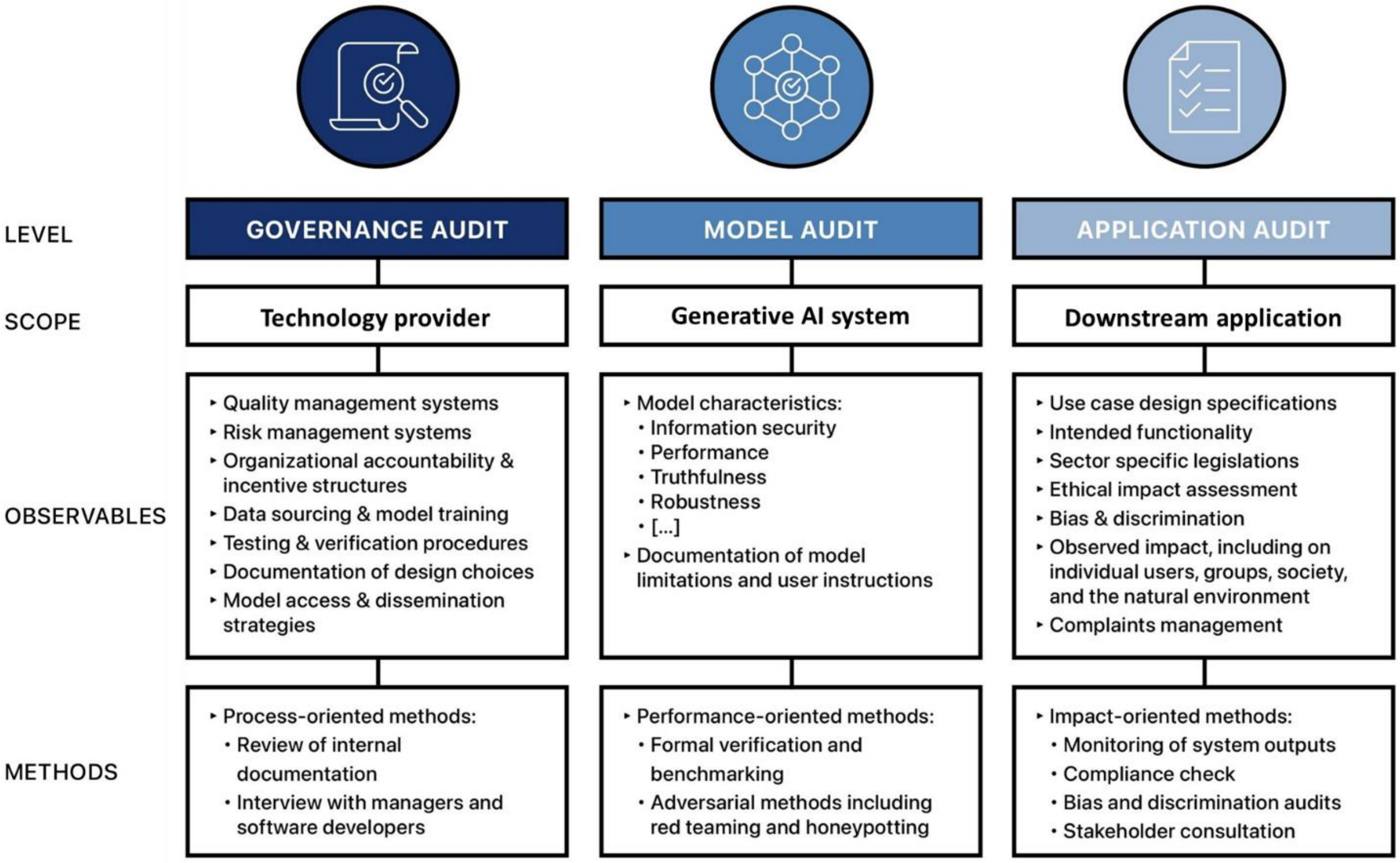

As illustrated in Figure 2, the scope and methodological affordances of *governance*, *model* and *application audits* differ in ways that make them critically complementary. Each of the three types of audits is individually necessary, yet they can only provide effective assurance when conducted in a structured and coordinated manner.[19] Further, their boundaries overlap and can be drawn in multiple ways. For example, governance audits are responsible for reviewing how training data is collected and pre-processed because it is a software development process. The characteristics assessed in model audits, however, may also reflect biases in training data, which may make it necessary to review data practices during model audits as well. Despite these overlaps, we believe the conceptual distinction between governance, model and application audits is useful for systematically approaching the risks posed by generative AI.

Of course, it would be theoretically possible to add further layers to our blueprint. For example, it would be conceivable to mandate governance audits of organisations developing downstream applications. However, we argue that the current combination of governance, model and application audits strikes an appropriate balance between covering a sufficiently large part of the development and deployment lifecycle, on the one hand, and being practically feasible to implement, on the other.

[19] N. Leveson, *Engineering a Safer World : Systems Thinking Applied to Safety*, Engineering Systems (MIT Press, 2011).

Regardless of how many layers are included, however, the success of our blueprint relies on responsible actors at each level who actively want to or are incentivised to ensure good governance.

Sections 4–6 will explain in detail what governance, model and application audits are, what they entail and how to conduct them. In the remainder of this section, we will discuss key takeaways from previous research and highlight the best practices that informed our three-layered approach. To do so, it is necessary to introduce and discuss distinctions and principles drawn from existing literature on AI auditing, IT auditing and systems engineering. To begin with, it is useful to make a distinction between *compliance* and *risk* audits. Compliance audits compare an entity's actions or properties to predefined benchmarks, requiring the promulgation of standards against which AI systems can be evaluated. Risk audits, by contrast, ask open-ended questions about how an AI system works to identify and mitigate risks. Mandating extensive compliance audits might seem appealing because it would streamline identifying and penalising the developers of harmful or illegal AI systems. However, the fact that compliance audits rely on the existence of a comprehensive set of standards makes them ill-equipped to handle generative AI systems. As discussed in Section 1, generative AI systems are adaptable to many downstream applications and can display emergent capabilities, which makes developing a set of standards that covers all or even most harms infeasible. As a result, our blueprint for how to audit generative AI systems incorporates elements of both risk audits (at the governance and model levels) and compliance audits (at the application level).

Further, a distinction can be made between *governance* and *technology* audits. Governance audits focus on the organisation designing or deploying AI systems. They evaluate software development processes, incentive structures, and allocation of roles and responsibilities. They are thus suitable for addressing risks best handled upstream but ill-equipped to handle downstream risks that only emerge over time as AI systems interact with complex environments. Technology audits are much better suited to those downstream risks. They focus on the technical properties of a system, including reviewing a model's architecture, checking its consistency with predefined specifications and repeatedly testing it to understand its workings and potential impact.[20] Since the best way to address generative AI risks depends on the type of risk (e.g., those concerning training data sourcing or those arising from application misuse), we argue that an effective auditing framework must include elements of both governance audits (upstream) and technology audits (downstream). Our blueprint satisfies this principle by conducting *governance audits* on the processes for building generative AI systems, *model audits* on the technical properties of pre-trained generative AI models and *application audits* on the technical properties of applications built on top of those pre-trained models.

Historically, technology audits have been applied to systems developed for specific functions in well-defined contexts (e.g., image analysis in radiology). Since generative AI systems can be widely adapted to many downstream applications, traditional technology audits will need to be updated to meet the specific governance challenges posed by generative AI. To do so, it is useful to distinguish between three different types of technology audits, namely: *functionality*, *model* and *impact audits*.[21] Functionality audits aim to understand the rationale behind AI systems by asking questions such as what is this system's purpose? Model audits review the system's decision-making logic. For symbolic AI systems,[22] this might entail reviewing source code, whereas, for sub-symbolic systems like generative AI, these audits look at how the model was designed, what data it was trained on and how it performs on different benchmarks. Finally, impact audits investigate the types, severity and prevalence of an AI system's effects on individuals, groups and the environment. These approaches are highly complementary to one another. However, because generative AI developers have limited information about how downstream developers and end-users might employ their systems, model audits will be especially important in communicating the intended uses and limitations of a pre-trained model. These audits will provide important context about a generative AI model's design and help mitigate unexpected downstream harms. Since governance audits and application audits are both already well-established practices in systems engineering and software development, this principle is a key justification for our inclusion of model audits and our blueprint's three-layered approach.

[20] D. Metaxa et al., "Auditing Algorithms: Understanding Algorithmic Systems from the Outside In," *Foundations and Trends in Human–Computer Interaction* 14, no. 4 (November 24, 2021): 272–344, https://doi.org/10.1561/1100000083.

Moving on, researchers and practitioners distinguish between *ex-ante* and *ex-post* technology audits. Ex-ante audits – which take place before a system is deployed – can pre-emptively identify and prevent some harms while informing downstream users about the model's appropriate, intended applications. Many of these ex-ante techniques, such as red teaming, already play an important role in auditing AI systems. However, these techniques cannot fully anticipate the risks from systems that continue to "learn",[23] a limitation that is especially relevant for generative AI systems, which can display unexpected emergent capabilities. Ex-post audits – which take place after a system is deployed – are better suited for systems that change over time. These audits can be further subdivided into *snapshot audits* (which occur once or on regular occasions) and *continuous audits* (which monitor performance over time). Most existing AI auditing procedures are snapshots,[24] but just as with ex-ante audits, snapshots are unable to provide meaningful assurances as generative AI systems display emergent capabilities or learn as they are fed new data. As a result, we advocate for the inclusion of ex-post, continuous audits at the application level of our blueprint for how to audit generative AI.

Finally, it is important to distinguish between *internal* and *external* audits. Whether an audit is internal or external depends on whether it is conducted by people within the audited organisation or by independent third parties. Internal audits have the advantage of privileged information access (e.g., access to information protected by privacy or IP rights), but without independent accountability, these audits struggle with outsiders perceiving self-reported findings as biased or untrustworthy and internal decision-makers ignoring recommendations that threaten business interests. These two disadvantages are especially pronounced for technologies with rapidly expanding capabilities and companies facing strong competition. Since both conditions apply to generative AI, we imagine the effectiveness of internal audits for this category of AI system will be limited. External audits, in contrast, address the concerns regarding biased evaluations and ignored recommendations but are constrained by limited access to internal processes.[25] Ultimately, we view trustworthiness and accountability as non-negotiable requirements of legally mandated audits and, therefore, advocate for external audits in our blueprint. To address the access issue, we turn to Koshiyama et al.'s typology, which distinguishes between different access levels.[26] At lower levels, auditors base their evaluations exclusively on publicly available information about the development process. At middle levels, auditors have access to the computational model itself, meaning they can manipulate its parameters and review its task objectives in their evaluations. Finally, at higher levels, auditors have access to all the details of a system, including full access to organisational processes, actual input and training data, and information about how and why the system was initially created. In Sections 4–6, we use this typology to indicate the level of access auditors must be granted to conduct audits at the governance, model and application levels.

[21] B. Mittelstadt, "Auditing for Transparency in Content Personalization Systems," *International Journal of Communication,* no 10, June (2016): 4991–5002.
[22] Symbolic AI systems are based on explicit methods like first-order logic and decision trees. Sub-symbolic systems rely on establishing correlations through statistical methods like Bayesian learning.
[23] V. Dignum, "Responsible Autonomy," in *Proceedings of the International Joint Conference on Autonomous Agents and Multiagent Systems, AAMAS*, vol. 1, 2017, 5, https://doi.org/10.24963/ijcai.2017/655.
[24] The post-market monitoring mandated by the EU AI Act is a rare example of continuous auditing.

To summarise, our blueprint incorporates elements of both compliance and risk audits, as well as governance and technology audits. Additionally, because generative AI systems are adaptable to a wide range of downstream tasks and can display emergent capabilities, our blueprint recommends technology audits include ex-post continuous evaluations and model-specific audits. Finally, to provide meaningful assurances about their reliability, all audits are conducted by independent third parties with varying levels of access to privileged information. In each of the next three sections, we describe what tasks the type of audit entails, the level of access it requires, and how they collectively help identify and mitigate the risks associated with generative AI systems.

## 4 Governance audits

At the first layer of our blueprint, technology providers developing generative AI systems should undergo governance audits that assess their organisational procedures, incentive structures and management systems. Governance audits have a long history in areas like IT governance and systems engineering,[27] where they ensure appropriate risk management systems are in place and promote procedural regularity and transparency throughout the software development lifecycle. They have also been shown to improve accountability by requiring companies to publish their results, preventing coverups and incentivising better behaviour. Specifically, we argue that governance audits of generative AI system providers should focus on at least the following three tasks:

1) *Reviewing the adequacy of organisational governance structures* to ensure that model development processes follow best practices and that quality management systems can capture risks specific to generative AI. While technology providers have in-house quality management experts, confirmation bias may prevent them from recognising critical flaws; involving external auditors addresses that issue. Nevertheless, governance audits are often most effective when auditors and technology providers collaborate to identify risks.[28] Therefore, it is important to distinguish accountability from blame at this stage of an audit.

2) *Creating an audit trail of the generative AI model development process* to provide documentation about the development of a model's capabilities, including information about its intended purpose, design specifications and choices, as well as training and testing procedures. Companies can use model cards,[29] datasheets,[30] and system cards[31] to create these audit trails. Specifying the purpose and capabilities of models upfront facilitates auditing and enforcement downstream (e.g., enforcing licensing agreements that only allow specific applications). Finally, requiring technology providers to document and justify their design choices should spark greater ethical deliberation by making trade-offs explicit.

3) *Mapping roles and responsibilities within organisations that design generative AI models* facilitates the allocation of accountability for system failures. The adaptability of generative AI models downstream does not absolve technology providers of all responsibility. Some risks are "reasonably foreseeable". In the related field of machine learning (ML) facial recognition, a study found that commercial gender classification systems were less accurate for darker-skinned females than lighter-skinned males.[32] After these findings were published, the technology providers quickly improved their models' accuracy, suggesting the main issue was inadequate risk management, rather than some inherent problem. Mapping the roles and responsibilities of different stakeholders can improve accountability and increase the likelihood of impact assessments being structured rather than ad-hoc, thus helping identify and mitigate harms proactively.

---

[25] M. Power, *The Audit Society: Rituals of Verification* (Oxford University Press, 1997)

[26] A. Koshiyama, E. Kazim, and P. Treleaven, "Algorithm Auditing: Managing the Legal, Ethical, and Technological Risks of Artificial Intelligence, Machine Learning, and Associated Algorithms," *IEEE* 55, no. 4 (2022): 40–50, https://doi.org/10.1109/MC.2021.3067225.

[27] S. Senft and F. Gallegos, *Information Technology Control and Audit*, 3rd ed. (Boca Raton: CRC Press, 2009).

---

[28] A. K. Chopra and M. P. SIngh, "Sociotechnical Systems and Ethics in the Large," in *Proceedings of the 2018 AAAI/ACM Conference on AI, Ethics, and Society*, AIES '18 (New York, NY, USA: Association for Computing Machinery, 2018), 48–53, https://doi.org/10.1145/3278721.3278740.

[29] M. Mitchell et al. 2019. "Model Cards for Model Reporting." In *Proceedings of the Conference on Fairness, Accountability, and Transparency* (ACM FAT* '19). New York, US, 220–229. https://doi.org/10.1145/3287560.3287596

[30] T. Gebru et al. 2021. "Datasheets for Datasets". *Communications of the ACM*. December 2021, Vol. 64 No. 12, Pages 86-92 DOI: 10.1145/3458723

[31] MetaAI, "System Cards, a New Resource for Understanding How AI Systems Work," 2023, https://ai.facebook.com/blog/system-cards-a-new-resource-for-understanding-how-ai-systems-work/. NOTE: Meta AI's detailed notes on the OPT model provide an exemplar of model training documentation.

[32] J. Buolamwini and T. Gebru, "Gender Shades: Intersectional Accuracy Disparities in Commercial Gender Classification," in *Proceedings of the 1st Conference on Fairness, Accountability and Transparency* (Conference on Fairness, Accountability and Transparency, PMLR, 2018), 77–91, https://proceedings.mlr.press/v81/buolamwini18a.html.

To conduct these three tasks, auditors will require the highest level of access (what Koshiyama et al. term white-box auditing). This includes knowledge of how and why a model was developed, as well as privileged access to facilities, documentation and personnel. It can even include access to the input data, learning procedures and task objectives used to train the model. This level of access may add to the logistical burden of governance audits by requiring nondisclosure and data-sharing agreements, but it is necessary for assessing generative AI models and the organisational safeguards around high-risk projects that providers may prefer not to discuss publicly. Importantly, this type of privileged access is not unique. It is standard practice in governance audits in other fields. IT auditors, for example, have full access to material and reports related to operational processes and performance metrics.[33]

The results of these governance audits should be tailored to the audience.[34] To the technology provider's management and directors, the auditors should provide a full report that directly and transparently discusses the vulnerabilities of existing governance structures. Such reports may recommend actions, but taking actions remains the provider's responsibility. Usually, such audit reports are not made public. Auditors should also create reports for two other audiences: regulators and developers of downstream applications. In some jurisdictions, hard legislation may demand that technology providers follow specific requirements. The EU AI Act, for example, requires providers to register high-risk AI systems with a centralised database.[35] In such cases, reports from independent governance audits can help ensure adherence to legislation. These reports can also inform downstream application developers' understanding of a model's intended purpose, testing, risks and limitations.

Finally, how do governance audits help identify and mitigate the potential harms of generative AI systems? Weidinger et al. listed six broad risk areas: discrimination, information hazards, misinformation hazards, malicious use, human-computer interaction harm, and automation and environmental harms. Governance audits address some of these directly. By assessing the adequacy of governance structures like structured access protocols for generative AI systems, these audits help reduce the risk of malicious use. They also reduce the risk of information hazards – which primarily stem from adversarial attacks that extract sensitive information from generative AI models – by reviewing the process by which datasets are sourced and the techniques by which models are trained. That said, we acknowledge governance audits only have an indirect impact (e.g., insofar as improved transparency reduces overall risks) on most of the risk areas listed by Weidinger et al. We find that the risk areas of discrimination, misinformation hazards and human-computer interaction harms are better addressed by model and application audits.

# 5 Model audits

At the second layer, we propose that generative AI systems should undergo model audits that assess their technical characteristics before they are adapted for downstream applications. These audits can provide critical insights about a model's capabilities and limitations, offering valuable information to both internal and external stakeholders.[36] For internal stakeholders, they can provide valuable benchmarks that inform model redesign and API licensing agreements, helping improve existing systems and limit unintended uses. Externally, clearly communicating the capabilities and limitations of a model can assist downstream developers in designing better applications that do not exceed a model's intended scope.

[33] Senft and Gallegos, *Information Technology Control and Audit*. (2018). New York: Auerbach Publicaitons.
[34] G. Falco et al., "Governing AI Safety through Independent Audits," *Nature Machine Intelligence* 3, no. 7 (2021): 566–71.
[35] European Commission, "The Artificial Intelligence Act," April 2021.

As mentioned in Section 1, evaluating a generative AI system independent of its applications is challenging. However, auditors can do so by leveraging two different approaches. The first seeks to identify generative AI systems' intrinsic characteristics. For example, the dataset on which a generative AI system has been trained can be assessed for completeness without reference to specific use cases.[37] However, for large datasets, this can be impractical or expensive.[38] The second approach involves testing a generative AI system's performance across multiple use cases and then aggregating the results. While that approach may be better suited for assessing generative AI systems' technical characteristics, it remains challenging to decide what use cases to include in the test and what model characteristics to test for. We recommend that model audits should test for characteristics that are

i. *socially and ethically relevant* (i.e., can be linked to specific risks);
ii. *predictably transferable* (i.e., impacts translate to downstream applications); and
iii. *meaningfully operationalisable* (i.e., can be assessed with available tools and methods).

With those criteria in mind, we suggest model audits should focus on (at least) the performance, information security and robustness of generative AI models. Other characteristics may meet the three criteria listed above, but we chose these three examples purely to illustrate how model audits can function in our framework.

1) *Performance,* i.e., how well the generative AI model functions on various tasks. Standardised benchmarks can help assess a model's performance by comparing it to a human baseline. For example, GLUE is a benchmark that aggregates a large language model's performance across multiple tasks into a single reportable metric.[39] The choice of tasks and benchmarks here is very important, with more generally being better. Some benchmarks have been criticised because the tasks are too narrow or easy and risk overestimating a model's capabilities. This likely happens because the model's performance on the task rapidly approaches that of a non-expert human, leaving little room for meaningful comparison.

[36] P. Adler et al., "Auditing Black-Box Models for Indirect Influence," *Knowledge and Information Systems* 54, no. 1 (January 1, 2018): 95–122, https://doi.org/10.1007/s10115-017-1116-3.
[37] I. D. Raji et al., "Outsider Oversight: Designing a Third Party Audit Ecosystem for AI Governance," in *AIES 2022 - Proceedings of the 2022 AAAI/ACM Conference on AI, Ethics, and Society*, 2022, 557–71, https://doi.org/10.1145/3514094.3534181.
[38] A. Paullada et al., "Data and Its (Dis)Contents: A Survey of Dataset Development and Use in Machine Learning Research," *Patterns* 2, no. 11 (November 12, 2021): 100336, https://doi.org/10.1016/j.patter.2021.100336.
[39] A. Wang et al. 2018. "GLUE: A Multi-Task Benchmark and Analysis Platform for Natural Language Understanding." In *Proceedings of the 2018 EMNLP Workshop BlackboxNLP*, pages 353–355, Brussels, Belgium. ACL. DOI: 10.18653/v1/W18-5446.

2) *Information security,* i.e., how difficult it is to extract training data from a generative AI model. Weidinger et al. identify "information hazards" as a category of risks in which AI models leak personal or sensitive information. Carlini et al. show that diffusion models like DALL-E 2 and StableDiffusion memorise images from the training data, which can then be successfully recovered by an adversarial actor attempting to extract that information.[40] Although differential privacy techniques may be able to reduce the amount of at-risk information, these methods are often too computationally expensive or severely degrade performance.[41] As such, auditing the extent to which models might expose information is important, though we will add the caveat that assessing a model's information security during model audits does not address all information hazards. In some cases, the risk of exposing sensitive information about users can only be determined (and audited) at the application level.

3) *Robustness,* i.e., how well the model reacts to unexpected prompts or edge cases. In ML, robustness indicates how well an algorithm performs when faced with new, potentially unexpected (i.e., out-of-domain) input data. A generative AI model that lacks robustness introduces the risks (among others) of critical system failures and adversarial attacks when it encounters data unlike what it has seen during training. As a result, researchers have created tools and methods for evaluating a model's robustness. Again, consider LLMs as an example. *AdvGLUE* is a benchmark that can be used to evaluate an LLM's susceptibility to adversarial attacks in different domains using a multi-task benchmark.[42] By quantifying robustness, AdvGLUE facilitates comparisons between LLMs. There are, however, many possible ways to operationalise robustness. Group robustness, for instance, measures a model's performance across different sub-populations.[43] To comprehensively assess different kinds of robustness, model audits should employ multiple tools and methods.

These three characteristics pertain to generative AI systems, but model audits should also review datasets as training models with biased or incomplete datasets can lead to poor robustness and discriminatory outcomes. Although curating "unbiased" datasets may be impossible, disclosing how a dataset was assembled can suggest its potential biases. Model auditors can then use existing tools and methods to interrogate biases in a pre-trained model. While the availability of such tools is encouraging, it is important to remain realistic about what these audits can achieve. They do not ensure generative AI systems are ethical in any global sense. Instead, they contribute to better precision in claims about a model's capabilities and inform the design of downstream applications.

---

[40] N. Carlini et al., "Extracting Training Data from Diffusion Models" (arXiv, January 30, 2023), https://doi.org/10.48550/arXiv.2301.13188.
[41] S. Lyu et al., "Differentially Private Latent Diffusion Models" (arXiv, May 25, 2023), https://doi.org/10.48550/arXiv.2305.15759.
[42] B. Wang et al., "Adversarial GLUE: A Multi-Task Benchmark for Robustness Evaluation of Language Models" (arXiv, January 10, 2022), https://doi.org/10.48550/arXiv.2111.02840.
[43] M. Zhang and C. Ré, "Contrastive Adapters for Foundation Model Group Robustness," in *ICML 2022 Workshop on Spurious Correlations*, 2022, https://doi.org/10.48550/arxiv.2207.07180.

Model audits require auditors to have privileged access to generative AI models and their training datasets. In the typology provided by Koshiyama et al., this corresponds to medium-level access, whereby auditors have access equivalent to a model's developer, meaning they can manipulate model parameters and review learning procedures and task objectives. This is required to assess generative AI systems' capabilities accurately during model audits. But in contrast to white-box audits, model auditors' access is limited to the technical system and does not extend to technology providers' organisational processes.

Some of the characteristics tested for during model audits correspond directly to the risks in Weidinger et al.'s taxonomy, such as information security and information hazards. However, it should be noted that our proposed model audits only focus on a few characteristics of generative AI systems. That is because the criterion of meaningful operationalisability sets a high bar: not all risks associated with generative AI can be addressed at the model level. Consider discrimination as an example. Model audits can expose how biases in models can be traced back to training datasets that reflect historic injustices. However, what constitutes unjust discrimination is often context-dependent and varies between jurisdictions, which makes it difficult to meaningfully address risks like unjust discrimination on a model level. This observation, however, does not argue against model audits but rather for complementary approaches like application audits, as discussed in the next section.

# 6 Application audits

In the third layer, we recommend that products and services built on top of generative AI models undergo application audits that assess the legality of their intended functions and how they will impact users and societies. Unlike governance and model audits, application audits focus on actors employing generative AI models in downstream applications.[44] Such audits are well-suited to ensure compliance with national and regional legislation, sector-specific standards and organisational ethics principles. We recommend application audits have two components: *functionality audits*, which evaluate applications using generative AI systems based on their intended and operational goals, and *impact audits*, which evaluate applications based on their impacts on different users, groups and the natural environment.

During *functionality audits*, auditors check whether an application's intended purpose is (1) legal and ethical in and of itself and (2) aligned with the intended uses of the underlying generative AI model. The first check assesses compliance with laws, regulations, ethics principles and codes of conduct relevant to the application. If an application is unlawful or unethical, its performance is irrelevant, and it should not be permitted on the market. The second check addresses risks stemming from how developers build on top of generative AI models, especially if they overstate or misrepresent what the application can do. To do so, these audits rely on the outputs of audits at other layers. For example, functionality audits can only assess whether developers' applications align with a model's intended use cases because governance

[44] B. Mittelstadt, "Auditing for Transparency in Content Personalization Systems." International Journal of Communication 10 (2016), 4991–5002

audits oblige technology providers to define those use cases. Similarly, functionality audits can only assess whether developers are overstating or misrepresenting their applications because model audits oblige technology providers to document limitations for downstream applications.

During *impact audits*, auditors disregard a system's design and intended purpose, focusing only on how it affects different users and the environment. The idea behind these audits is simple: every system can be understood exclusively in terms of its inputs and outputs. But implementing them is often easier said than done. AI systems evolve with their environments in unexpected ways, which can make identifying and tracking an application's indirect, second-order impacts incredibly challenging, as there may be no clear link between a model's intended purpose and the system's actual impact. Overcoming this challenge may require legislators to clearly define which direct and indirect impacts are legally and socially relevant on a case-by-case basis. Even then, attributing causality to a single system in a complex environment can be even more difficult (e.g., consider how much researchers have struggled to determine whether Facebook's content recommendation algorithms cause polarisation). As such, we recommend applications undergo both pre-deployment (ex-ante) and post-deployment (ex-post) evaluations.

Pre-deployment evaluations can leverage empirical and analytical strategies to assess applications before they are released. For example, *sandbox environments* can allow developers and policymakers to understand an application's potential impact by providing a sandbox in which to identify and mitigate biases.[45] Real-world environments, however, may differ from training and testing in unexpected ways, which is why it is important to also include analytical assessment strategies like *ethical foresight analysis*.[46] Post-deployment monitoring can also be done in different ways, such as through automated oversight programs that continuously monitor systems for impermissible outputs or through more tailored evaluations by private companies or government agencies. Overall, application audits seek to ensure that ex-ante testing and impact assessments have been conducted following existing best practices, that there are post-market plans to ensure continuous monitoring of system outputs, and that procedures are in place to mitigate or report different types of failure modes.

Lower levels of access are typically sufficient for application audits. To assess an application for legal compliance or to quantitatively assess the relationship between a system's inputs and outputs, auditors do not need especially privileged access. Standard API access to the generative AI application should be enough, though in some cases, auditors may also need access to input data as well. Koshiyama et al. term these lower levels of access black-box model access and input data access, respectively.

Application audits are well-suited to address risks that are more context-dependent, such discrimination and human-computer interaction harms. Methodologically, application audits can address these harms using quantitative assessments, which give a broad sense of the types of outputs an application produces, and qualitative assessments, which more deeply consider users' lived experiences interacting with the applications. For example, a generative AI model that performs poorly for some social groups may permit unjust discrimination at the application level, creating representational harms through inaccurate portrayals or allocational harms through its use in important decisions like hiring. Our main point is that by focusing on individual use cases, application audits, specifically the functionality and impact audits described above, are the best way to mitigate harms that require context to understand. That said, we do want to re-emphasise that application audits may struggle to address different forms of harm in any global sense. The interpretation of toxicity and discrimination can vary greatly across cultural, social or political groups, and so what is acceptable in one setting may not be in others.

[45] J. Truby et al., "A Sandbox Approach to Regulating High-Risk Artificial Intelligence Applications," *European Journal of Risk Regulation* 13, no. 2 (June 2022): 270–94, https://doi.org/10.1017/err.2021.52.

[46] L. Floridi and A. Strait, "Ethical Foresight Analysis: What It Is and Why It Is Needed?," *Minds Mach (Dordr)* 30, no. 1 (2020): 77–97, https://doi.org/10.1007/s11023-020-09521-y.

Although we believe all technology providers designing and disseminating generative AI models should be required to undergo governance audits and model audits, we recommend that application audits be employed more selectively. Importantly, we do not want these application audits to be conflated with certification. Although application audits may form the basis for certification, certification is different in that it requires separate, predefined standards against which a product or service can be evaluated and institutions that can ensure the certification process's integrity. Regardless, the results of application audits should be made publicly and easily accessible to cure information asymmetries in tech policy and incentivise companies to correct behaviour.

## 7 Clarifications and limitations

In this section, we tie together loose ends from the previous three sections and discuss the limitations of our approach. On the first point, we illustrate how audits on the three different layers work together and how one might select independent, third-party auditors. On the second, we discuss four limitations of auditing approaches to governing generative AI: one conceptual, one institutional, one technical, and one practical.[47] First, all audits will struggle with operationalising normative concepts like robustness. Second, we currently lack an institutional ecosystem that can support independent third-party audits of generative AI. Third, static auditing procedures will quickly become dated as technological innovation accelerate and generative AI systems operate with ever higher degrees of autonomy, complexity, and adaptability. Finally, not all risks from generative AI can be practically addressed on the technology level.

To adequately address the multiplicity and complexity of risks from generative AI, we have argued that governance, model and application audits must work together in a complementary, structured process. In practice, this means that outputs from audits on one level become inputs for audits on other levels. In our blueprint, for instance, model audits produce reports summarising a model's properties and limitations, which should inform application audits that verify whether a model's known limitations have been considered when designing downstream applications. Similarly, ex-post application audits produce output logs documenting the impact that different applications have in applied settings. Such logs should inform the underlying model's continuous redesign and revisions to accompanying model cards. Finally, governance audits must check the extent to which technology providers' software development processes and quality management systems include mechanisms to incorporate feedback from application audits. Figure 3 illustrates how governance, model and application audits are interconnected in our blueprint.

[47] J. Mökander et al., "Ethics-Based Auditing of Automated Decision-Making Systems: Nature, Scope, and Limitations," *Science and Engineering Ethics* 27, no. 4 (July 6, 2021): 44, https://doi.org/10.1007/s11948-021-00319-4.

*Figure 3. Outputs from audits on one level become inputs for audits on other levels.*

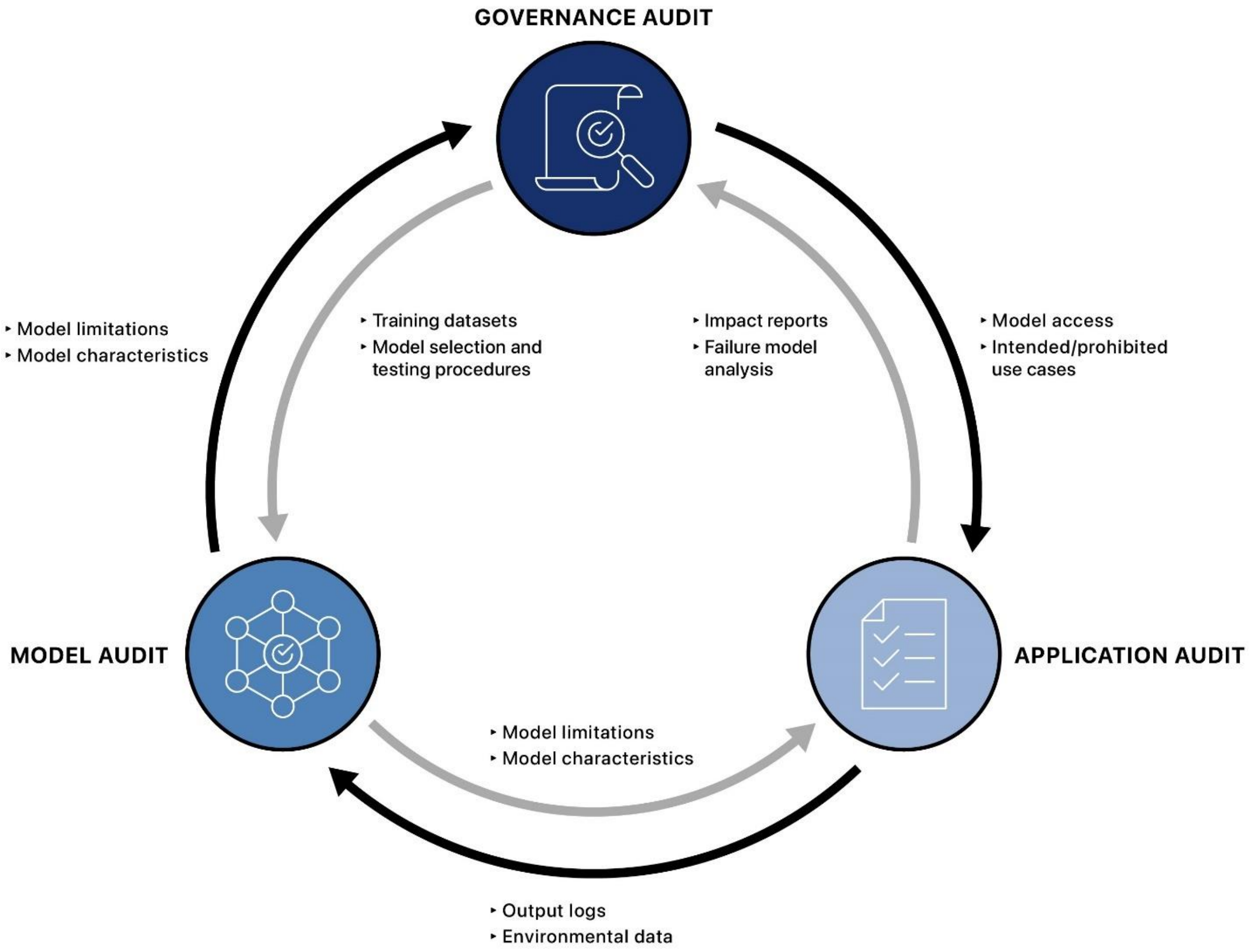

How to select independent third-party auditors is also an important question because each layer in our auditing blueprint requires them. First, the same organisation need not conduct all three types of audits as each requires different skills. Governance audits may require an understanding of corporate governance and soft skills like stakeholder communication, while model audits are highly technical and application audits typically require domain-specific expertise. All these competencies may not be found within one organisation. Second, institutional arrangements may vary between sectors and jurisdictions, so the best option for performing various elements of governance, model or application audits may depend on the specific institutions within a geography or industry. For example, since medical devices are already subject to various testing and certification procedures, one could imagine incorporating application audits for new medical devices that use generative AI into these existing procedures. The main points are that different independent auditors can perform the three different types of audits and different institutional arrangements may be preferable in different jurisdictions or sectors.

We now move on to limitations. In Sections 5 and 6, we discussed how audits may struggle to operationalise or create standard evaluation metrics for normative concepts like robustness and toxicity. This problem exists with all attempts to develop effective AI auditing procedures. Many normative concepts are not only context-dependent (i.e., different cultures view toxicity differently), but they are also hard to quantify (i.e., what does 7/10 robustness mean?). Additionally, once a measure is widely adopted for auditing generative AI systems, companies may be able to game it by over-indexing on that measure. As Goodhart's Law reminds us, "When a measure becomes a target, it ceases to be a good measure." Creating a single set of perfect evaluation metrics may end up being impossible, but an important avenue for further research is developing new and better methods of operationalising normative concepts in ways that are verifiable and resilient to manipulation.

A second important limitation is the lack of an institutional auditing ecosystem, especially since an auditing procedure is often only as good as the institutions conducting it. We leave this question open in our blueprint for two reasons: first, to provide flexibility to different institutional ecosystems that vary across jurisdictions and sectors; second, because our blueprint is flexible enough to be adopted by any external auditors. That said, we will briefly discuss what can be learned from existing auditing practices below. There are five main institutional arrangements:

1) *Private service providers*, chosen by and paid for by the technology provider.
2) *A government agency,* centrally administered and paid for by government, industry or a combination of both.
3) *An industry body*, operationally independent yet funded through fees from its member companies.
4) *Non-profit organisations*, operationally independent and funded through public grants and voluntary donations.
5) *An international organisation*, administered and funded by its member countries.

Each of these arrangements has its own set of advantages and disadvantages. For example, private service providers are incentivised to innovate to win business, helping them keep pace with the fast-moving field of generative AI. But they are also reliant on good relationships with the companies they audit for their business, increasing the risk of collusion. We recommend further research to investigate the effectiveness of different institutional arrangements for conducting and enforcing the three types of audits proposed.

A third limitation is that any static auditing procedure risks becoming obsolete as technological innovation accelerates. Generative AI systems differ in autonomy, complexity, and adaptability. Hence, procedures well-suited to audit a specific system may fail to account for the governance challenges posed by another. For example, it is hard to predict the failure modes of generative AI systems that operate with a high degree of autonomy in dynamic environments. This makes their formal specification and verification uniquely challenging. They point is that future generative AI systems are likely to become ever more autonomous, complex, and adaptable. As technologies evolve, so too must the auditing procedures put in place to ensure good governance.

A final limitation is that not all risks associated with generative AI can be addressed on the technology level. In the previous sections, we highlighted how our three-layered approach can help identify and mitigate example risks from Weidinger et al.'s taxonomy. The cases we described, however, are illustrative, not exhaustive. Consider automation harm as an example. Improving generative AI capabilities may automate away the need for certain jobs like translators, copywriters or illustrators. This possibility raises important concerns about how automation might impact people's livelihoods and how the fruits of automation are distributed in society. In these cases, it is important to remain realistic about what audits can do and not overpromise when introducing new governance mechanisms. Many automation harms are unlikely to be addressed through audits and will likely require social and political reforms.[48] We think this provides another direction for further research: how can we create social and political reforms that complement technically oriented mechanisms like audits in a way that allows for better holistic governance of generative AI?

# 8 Conclusion

Existing governance mechanisms, including past efforts to audit AI, fall short with generative AI systems that are adaptable to a wide range of downstream applications. In this chapter, we have attempted to bridge this governance gap by outlining a blueprint for how to audit generative AI systems. Specifically, we have advocated for a three-layered approach, whereby governance, model and application audits inform and complement each other.

During *governance audits*, technology providers developing generative AI systems undergo assessments of their organisational procedures, accountability structures and quality management systems. During *model audits*, a model's capabilities and limitations are assessed along several dimensions, including performance, information security and robustness. *During application audits*, products and services built on top of generative AI models undergo functionality and impact audits that assess the legality of their intended functions and their potential impact on users.

It is important to remember that the effectiveness of this auditing approach derives from the way it combines and coordinates different types of audits. Governance audits should ensure that providers have mechanisms to take the output logs generated during application audits into account when redesigning their models. Similarly, application audits should ensure that downstream developers take the limitations identified during model audits into account when building on top of a specific model.

[48] J. Mökander & R. Schroeder. "Artificial Intelligence, Rationalization, and the Limits of Control in the Public Sector: The case of Tax Policy Optimization", *Social Science Computer Review* (2024): 1-20, https://doi.org/10.1177/08944393241235175

However, even under ideal circumstances, audits will not protect against all the risks associated with generative AI. We remind readers of the main limitations that we discussed in Section 7. Namely, the difficulty of operationalising normative concepts, the lack of an institutional auditing ecosystem and the fact that not all harms can be addressed at the technology level.

This leaves us with one last question: what are the implications for lawmakers and technology providers? First, lawmakers can help develop an institutional ecosystem for conducting and enforcing governance, model, and application audits of generative AI systems. For example, they can encourage private sector auditing initiatives by developing standardising evaluation metrics or rewarding achievements through monetary incentives.[49] They can also draw on this three-layered blueprint to develop clearer, more operationalisable systems requirements. In doing so, they will encourage companies to invest in cost-effective, comprehensive auditing tools. Fortunately, no one needs to start from scratch. Technology providers and lawmakers are already experimenting with some of the auditing activities we mentioned. Consequently, when thinking about mandating governance, model and application audits, lawmakers and regulators can leverage a wide range of existing tools and methods, including impact assessments, benchmarking, model evaluation and red teaming.

Technology providers can also get ahead of the curve. To start with, they can subject themselves and their models to governance and model audits, simultaneously creating demand for independent auditing bodies and helping spark methodological innovation as these procedures are used in practice. In the medium term, providers can demand that products and services built on top of their models undergo application audits, enforcing this requirement through structured access procedures that make use conditional on these audits. Although this requirement may risk losing some downstream applications, requiring application audits can help technology providers build a better brand and minimise future financial and legal risks. Finally, in the long term, technology providers should establish and fund an independent industry body that conducts or commissions governance, model and application audits.

It is worth concluding this chapter with some words of caution. First, we recognise that future developments may change the feasibility and effectiveness of this blueprint. For example, governance audits make sense when a limited number of actors have the resources and talent to train and disseminate generative AI models. If this process is democratised – either through reduced barriers-to-entry or a shift to open-source everything – then the form of, and potentially the need for, governance models will need to be reassessed. As a result, while maintaining the usefulness of our three-layered approach, we acknowledge that it will need to be continuously revised in response to the changing technological and regulatory landscape. Finally, our blueprint is intended to complement and interlink existing governance models by strengthening procedural transparency and regularity. Instead of being adopted wholesale by lawmakers and technology providers, we hope our three-layered approach will be adopted, adjusted and expanded to meet the governance needs of different stakeholders and contexts.

---

[49] L. Floridi et al., "AI4People-An Ethical Framework for a Good AI Society: Opportunities, Risks, Principles, and Recommendations," *Minds and Machines* 28, no. 4 (2018): 689–707, https://doi.org/10.1007/s11023-018-9482-5.